\documentclass[lettersize,journal]{IEEEtran}
\usepackage{amsmath,amsfonts}
\usepackage{algorithmic}
\usepackage{algorithm}
\usepackage{array}
\usepackage[caption=false,font=normalsize,labelfont=sf,textfont=sf]{subfig}
\usepackage{textcomp}
\usepackage{stfloats}
\usepackage{url}
\usepackage{verbatim}
\usepackage{graphicx}
\usepackage{cite}
\usepackage{placeins}
\usepackage{gensymb}
\usepackage{xcolor}
\begin{document}

\title{\textit{EV-EcoSim}: A grid-aware co-simulation platform for the design and optimization of electric vehicle charging infrastructure}

\author{Emmanuel Balogun\IEEEauthorrefmark{1} \textit{Student Member, IEEE}, Elizabeth Buechler\IEEEauthorrefmark{1} \textit{Student Member, IEEE}, Siddharth Bhela\IEEEauthorrefmark{2} \textit{Member, IEEE}, Simona Onori\IEEEauthorrefmark{3} \textit{Senior Member, IEEE}, and Ram Rajagopal\IEEEauthorrefmark{4} \textit{Member, IEEE}
\thanks{{ \IEEEauthorblockA{\IEEEauthorrefmark{1}Department of Mechanical Engineering}, \IEEEauthorblockA{\IEEEauthorrefmark{3}Department of Energy Science and Engineering}, and \IEEEauthorblockA{\IEEEauthorrefmark{4}Department of Civil and Environmental Engineering and Department of Electrical Engineering, Stanford University, Stanford, CA, USA}. \IEEEauthorblockA{\IEEEauthorrefmark{2}Siemens Technology, Princeton, NJ, USA}.
    E. Balogun was supported by the Chevron Energy Fellowship and Stanford Bits and Watts. All accompanying software is cited in the contributions of this paper.}
    }
\thanks{}}

\maketitle

\begin{abstract}
To enable the electrification of transportation systems, it is important to understand how technologies such as grid storage, solar photovoltaic systems, and control strategies can aid the deployment of electric vehicle charging at scale. In this work, we present \textit{EV-EcoSim}, a co-simulation platform that couples electric vehicle charging, battery systems, solar photovoltaic systems, grid transformers, control strategies, and power distribution systems, to perform cost quantification and analyze the impacts of electric vehicle charging on the grid. This python-based platform can run a receding horizon control scheme for real-time operation and a one-shot control scheme for planning problems, with multi-timescale dynamics for different systems to simulate realistic scenarios. We demonstrate the utility of \textit{EV-EcoSim} through a case study focused on economic evaluation of battery size to reduce electricity costs while considering impacts of fast charging on the power distribution grid. We present qualitative and quantitative evaluations on the battery size in tabulated results. The tabulated results delineate the trade-offs between candidate battery sizing solutions, providing comprehensive insights for decision-making under uncertainty. Additionally, we demonstrate the implications of the battery controller model fidelity on the system costs and show that the fidelity of the battery controller can completely change decisions made when planning an electric vehicle charging site.
\end{abstract}


\section{Introduction}
\IEEEPARstart{A}{s} the electrification of the transportation sector continues to accelerate, it is increasingly important to understand the interactions between transportation infrastructure, power systems, and consumer behavior. Rapid deployment of electric vehicles (EVs) is projected to persist, approaching 20 million by 2030 \cite{na_website_nda}. A major deterrent to EV adoption is range anxiety, which can be alleviated by fast charging. The increasing deployment of EV fast charging stations will have significant impacts on power distribution systems by increasing voltage violations and accelerating transformer aging \cite{borlaug_2021}.

To alleviate grid impacts, many studies have suggested pairing EV chargers with battery energy storage systems (BESS) and other distributed energy resources (DERs) such as solar photovoltaic (PV).  If sized properly, BESSs can be leveraged for energy arbitrage and frequency regulation, which can reduce electricity bills and make revenue in the electricity markets to pay back its cost over its lifetime \cite{brivio_2016, hesse_2017, hesse_2019}. Evaluating the optimal sizing, system design, and operation of those resources requires consideration of the entire EV ecosystem, including the EV charging behavior, DERs, distribution grid, transformers, and other resources. Each of these resources has dynamics that operate at different timescales. Interactions between these subsystems can be modeled through co-simulation platforms that tie together these dynamics with multi-timescale and multi-fidelity simulation. 

Modelling the entire EV ecosystem is also important for understanding the tradeoffs between objectives for different stakeholders.  For example, utilities may want to minimize voltage violations and defer distribution circuit and power transformer upgrades, while EV charging companies may want to maximize their station reliability and profit. Depending on who owns grid storage, maximizing its operational value while minimizing its degradation may be important.

An example of a platform that can include EV charging is HELICS \cite{palmintier_2017}, which integrates power transmission and distribution infrastructure. HELICS can be useful for large-scale co-simulation, however, it may not be immediately suited to some Electric Vehicle Supply Equipment (EVSE) operation and planning problems. It will require that the key components of interest, such as EV charging, batteries, and transformers be developed and integrated into the platform, which can be arduous. Authors in \cite{panossian_2023} use HELICS to investigate the impacts of EV charging at scale for the San Francisco Bay Area region. However, because major charging bottlenecks occur at the distribution level \cite{borlaug_2021}, the impacts of EVSE on local networks is critical for planning and operation. Additionally, the heterogeneity of distribution circuits make them critical to understanding the impact of EV charging on local community grid resilience and equity \cite{brockway2021inequitable}. Another tool \cite{xue2020simulator}, was introduced by the World Resources Institute, for which authors claimed could be used to estimate and manage EV load impacts on low-voltage distribution grids. However, the work is focused on EV charging load profiles, similar to \cite{powell_2022}. It does not incorporate the battery and transformer dynamics nor aging, and there is no power flow simulation for grid-integration analyses. In another work \cite{sarieddine2023real}, the authors proposed a real-time co-simulation test-bed to investigate cybersecurity and communication within the EV ecosystem.

Co-simulators are commonly developed for a specific theme of problems. Notably, there is a dearth of works that consider EV infrastructure planning with detailed distributed energy resource (DER) models, grid-integration, and cost analyses.

When modelling EV charging infrastructure, it is also important to evaluate how the fidelity of subsystem models affects results and computational tractability. Literature on the optimization of DERs for planning and operations of EV charging generally use optimization formulations with simplified linear models, particularly for the battery systems \cite{hesse_2017a, hesse_2019, negarestani_2016, liu_2020, salapic_2018}. While nonlinear models are often more accurate, they are generally not used because of the computational complexity and the data and effort needed for validation. However, some works have suggested that the use of higher fidelity battery models can enable greater economic value over operational lifetimes \cite{reniers_2018, onori2019optimizing, reniers_2021}. Understanding how model fidelity, particularly for battery systems, affects design decisions for the operation and planning of EV charging stations is still an open area of research.

In this work, we address these research gaps by developing \textit{EV-EcoSim} \cite{Balogun_EV-EcoSim_A_grid-aware_2023}, a python-based multi-timescale co-simulation platform that integrates EV charging, battery system identification and degradation, power systems, and controls, to value pathways for scaling EV fast-charging using DERs. The modules within our tool can also be adapted to other simulators, if desired. This platform enables scenario-based evaluation of different designs over the lifetime of the DER-integrated EVSE using high-fidelity models of system components. Different solar, battery, and transformer sizes can be evaluated under different optimization objectives and the expected costs/revenue and grid impact of different objectives at varying EVSE utilization levels can be calculated. Co-simulation with the power distribution system enables evaluation of charging impacts on network voltages and transformer aging. The main contributions of this paper are as follows:
\begin{enumerate}
    \item The development of the \textit{EV-EcoSim} platform. This platform serves as a valuable tool for orchestrating operations and planning involving diverse stakeholders, including electric utilities, EV service providers, and agencies involved in energy transition planning. To the best of our knowledge, this is the first (python-based) open-source platform that allows end-to-end simulation and optimization of grid-integrated DERs and EVSE systems. Notably, it encompasses comprehensive assessments of both cost implications and grid impact of EVSEs. Furthermore, in this paper, we present a practical application of \textit{EV-EcoSim} by demonstrating its utility in sizing a collocated battery for an EV charging station in conjunction with a fixed solar PV capacity. This demonstration simplifies an otherwise intricate evaluation of system economics and operational ramifications on the local power distribution network.
    \item Within this platform, we provide the option of a receding horizon control (RHC) scheme, which can be used to study real-time operation of an EVSE and a one-shot control scheme, which is suitable for planning EV charging infrastructure.
    \item We develop a battery system identification module with an open-circuit voltage correction scheme, which can be used to calibrate battery models from available experimental data. The method significantly improves the equivalent circuit battery (ECM) model, with Mean Absolute Percent Errors (MAPE) around 0.25\%. Additionally, we include logic for ensuring that the cells in the battery system do not reach unsafe voltages, which is a detail that many battery optimization studies ignore. 
    \item We demonstrate the implications of the battery controller model fidelity on the system costs and show that the fidelity of the battery controller can completely alter decision-making when planning an electric vehicle charging site, implying that the incongruence between the model assumptions and the true system should be taken into account during planning.
\end{enumerate}

The rest of the paper is organized as follows. Section II includes a description of the simulation framework, including components that make up the system and how the states evolve in time. In Section III, a case-study that leverages \textit{EV-EcoSim} is introduced. In Section IV we discuss the results and in Section V conclusions and comments on future work are provided.

\section{Simulation framework}
\begin{figure*}
    \centering
    \includegraphics[width=\linewidth]{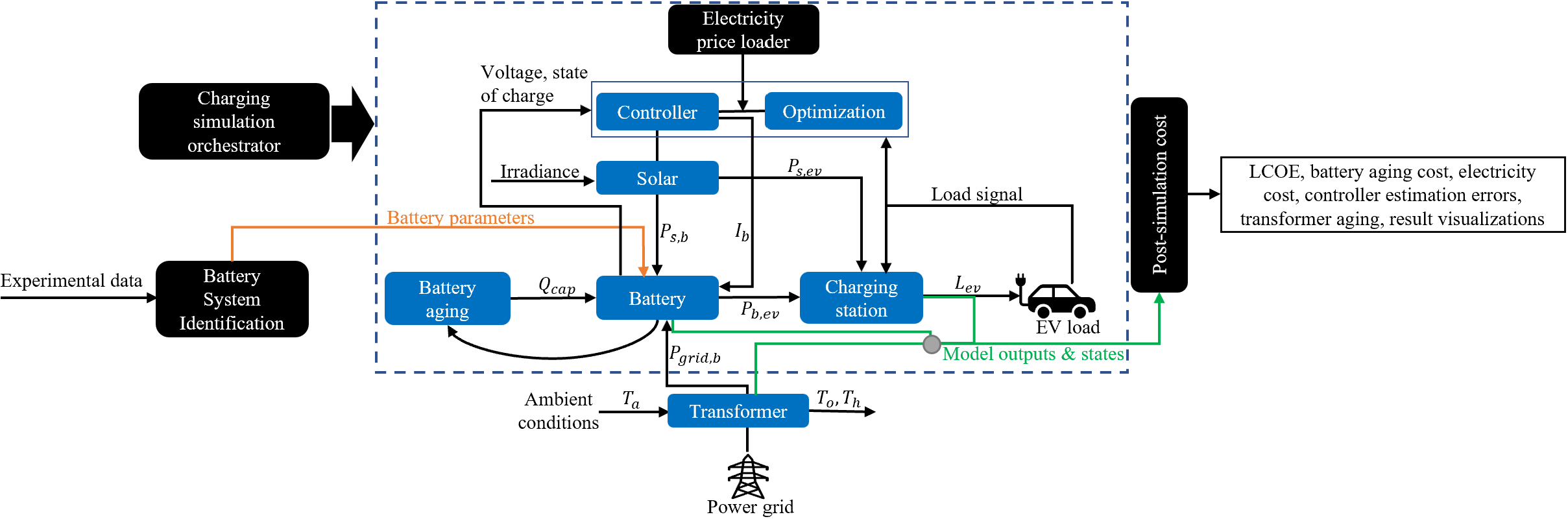}
    \caption{System diagram for simulation framework showing the various objects and flow of inputs/outputs.}
    \label{fig:sim_frame}    
\end{figure*}

Fig. \ref{fig:sim_frame} describes the \textit{EV-EcoSim} framework. Every subsystem is built as a module. We emulate physical systems using their respective dynamical models and their interactions are captured through data exchange at each time step. The framework includes a controller that can receive feedback from simulated physical systems. The charging simulation orchestrator initializes all modules (within the dashed region) using their respective configurations and populates all charging station modules at specific nodes in the power network. 

The optimization and controller modules exist separately by design but work in unison. The controller loads the optimization module, takes as input the load signal from the load generator, and sets up the problem to be solved. Afterwards, the controller ingests solutions from the optimization module and sends control signals to the solar and battery modules. All desired time series for the module states are saved during and after simulation. The post-simulation cost module produces estimates for system lifetime costs and grid impact. $P_{b, ev}$ and $P_{s, ev}$ are the powers from the battery and solar PV system to the EVSE, respectively. $I_b$ is the current signal from the controller to the battery. $P_{s,b}$ is the power from solar to battery. $P_{grid,b}$ is the power from the grid to the battery. $Q_{cap}$ is the battery's capacity. $T_a$, $T_o$, and $T_h$ are the ambient, transformer top-oil, and transformer hot-spot temperatures, respectively. The individual modules are discussed within the rest of this section.

\subsection{Battery Dynamics}
Batteries constitute a significant portion of the cost in electrification projects, accounting for about 20-30\% of the overall cost of a passenger EV \cite{na_website_nde}. Using an accurate battery model is important to maximize its value. Modelling approaches today include bucket, physics-based, and equivalent circuits.
 \begin{enumerate}
     \item \textit{Bucket Models (BM)}: BMs are the simplest and most common battery models used in optimization studies \cite{hesse_2017, negarestani_2016, liu_2020, salapic_2018, yan_2019}. They ignore battery physics and treat the battery as an energy container from which power can be consumed at a constant rate. 
     \item \textit{Physics-based models (PBM)}: PBMs are electrochemical models developed from first principles\cite{liu2019aging}. PBMs model the internal components of the battery cell and range in complexity in the details of the components they capture. Examples of physics-based models include the single particle model (SPM), pseudo-2D model (P2D).
     \item \textit{Equivalent Circuit Model (ECM)}: Due to their computational tractability, ECMs are often used in battery management systems \cite{ahmed2015model}. They are data-driven semi-empirical models that use a circuit model to capture battery dynamics, rather than modelling the physics of the battery's internal components. A well-calibrated ECM with the right context may outperform PBMs. 
 \end{enumerate}

\begin{figure}
    \centering
    \includegraphics[width=7cm]{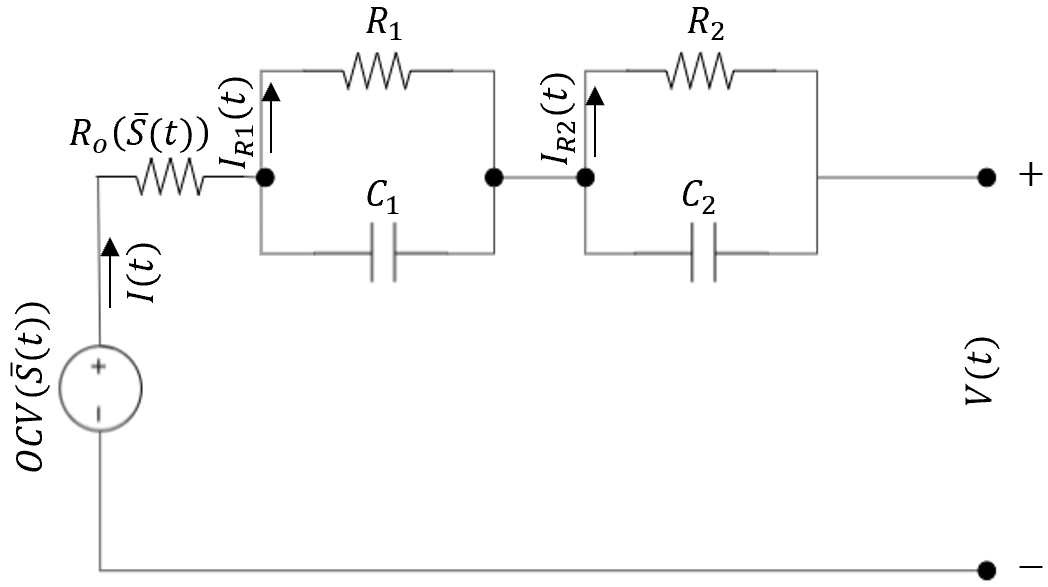}
    \caption{Second order ECM (Thevenin) circuit diagram for the cell}.
    \label{ECM}
\end{figure}

We elect ECMs due to their modest trade-off between computational complexity and accuracy. The ECM consists of a single resistance $R_o$ and two resistance-capacitor (RC) pairs; see Fig. \ref{ECM}. A cell is modelled and scaled by combining the cells in series and parallel to form a pack. Pack configurations can be specified by the user by setting the desired voltage and energy capacity. It is assumed the cells are maintained at constant ambient conditions at $23\degree C$ (via an auxiliary cooling system), which is the temperature at which the battery experiments were carried out.  Modelling at higher ambient temperatures will require experiments for each desired temperature and may quickly become prohibitively expensive. However, as more data become available, model parameters at different temperatures can be included within the battery module. The battery's internal thermal states are not explicitly modelled in this paper and will be explored in future work.

$OCV(\bar{S}(t))$ is the open circuit voltage of the cell and a function of the battery cell’s state of charge (SoC) $\bar{S}(t)$ at time $t$. The voltage equation for the RC-circuit in Fig. \ref{ECM} is:
\begin{equation}\label{eq1}
    V(t) = OCV(\bar{S}(t)) - I(t)R_o(\bar{S}(t)) - I_{R_1}(t)R_1 - I_{R_2}(t)R_2,
\end{equation}
stating that the voltage across the terminals of the cell is equivalent to the open circuit voltage minus the voltage drop across the resistive components of the cell. The current flowing through $R_o$, $I(t)$ is the current induced by the load to which the battery is connected to, or the discharge or charge current of the battery. The power flowing out (-) or into (+) the battery at time $t$ is:
 \begin{equation}
     P(t) = V(t)I(t) .
 \end{equation}
 Each RC pair comprises a resistor and capacitor that are connected in parallel, thus the sum of the current flowing through the resistor and capacitor must be equal to the current $I(t)$ flowing into the connecting node:
\begin{equation}\label{eq3}
    I(t) = I_{R_1}(t) + C_1\frac{dV_{C_1}}{dt}
\end{equation}
\begin{equation}\label{eq4}
    I_{R_1}(t) = I(t) - R_1C_1\frac{dI_{R_1}}{dt}
\end{equation}
where ${dV_{C_1}}/{dt}$ is the rate of voltage change across capacitor $C_1$ and ${dI_{R_1}}/{dt}$ is the rate of current change across resistor $R_1$. Equations \eqref{eq3} and \eqref{eq4} hold similarly for the second RC pair R2-C2, by simply replacing the indices. These relationships hold if one desires to increase the order of the ECM by including more parallel RC pairs. A more detailed treatment of battery state evolution can be found in \cite{plett2015battery}.

The following section describes how experimental data can be used to identify realistic estimates for $R_o, R_1, C_1, R_2,$ and $C_2$ which can then be used in simulation. If an \textit{EV-EcoSim} user has experimental data for the battery system they wish to simulate, this methodology can be used to fit a model.

\subsection{\textit{Battery System Identification}}
The Battery System Identification module is important for accurate representation of the battery dynamics. We demonstrate the system identification module on data from 3 of 10 identical Nickel Manganese Cobalt (NMC) cells, each subjected to Urban Dynamometer Driving Schedule (UDDS) cycles, Hybrid Pulse Power Characterization (HPPC), and Electrochemical Impedance Spectroscopy (EIS) diagnostic tests\cite{pozzato_2022}. The cells were tested at different charging rates, ranging from C/4 to 3C, where 1C rate is equivalent to discharging the entire battery cell capacity in 1 hour.

\begin{figure}
    \centering
    \includegraphics[width=7cm]{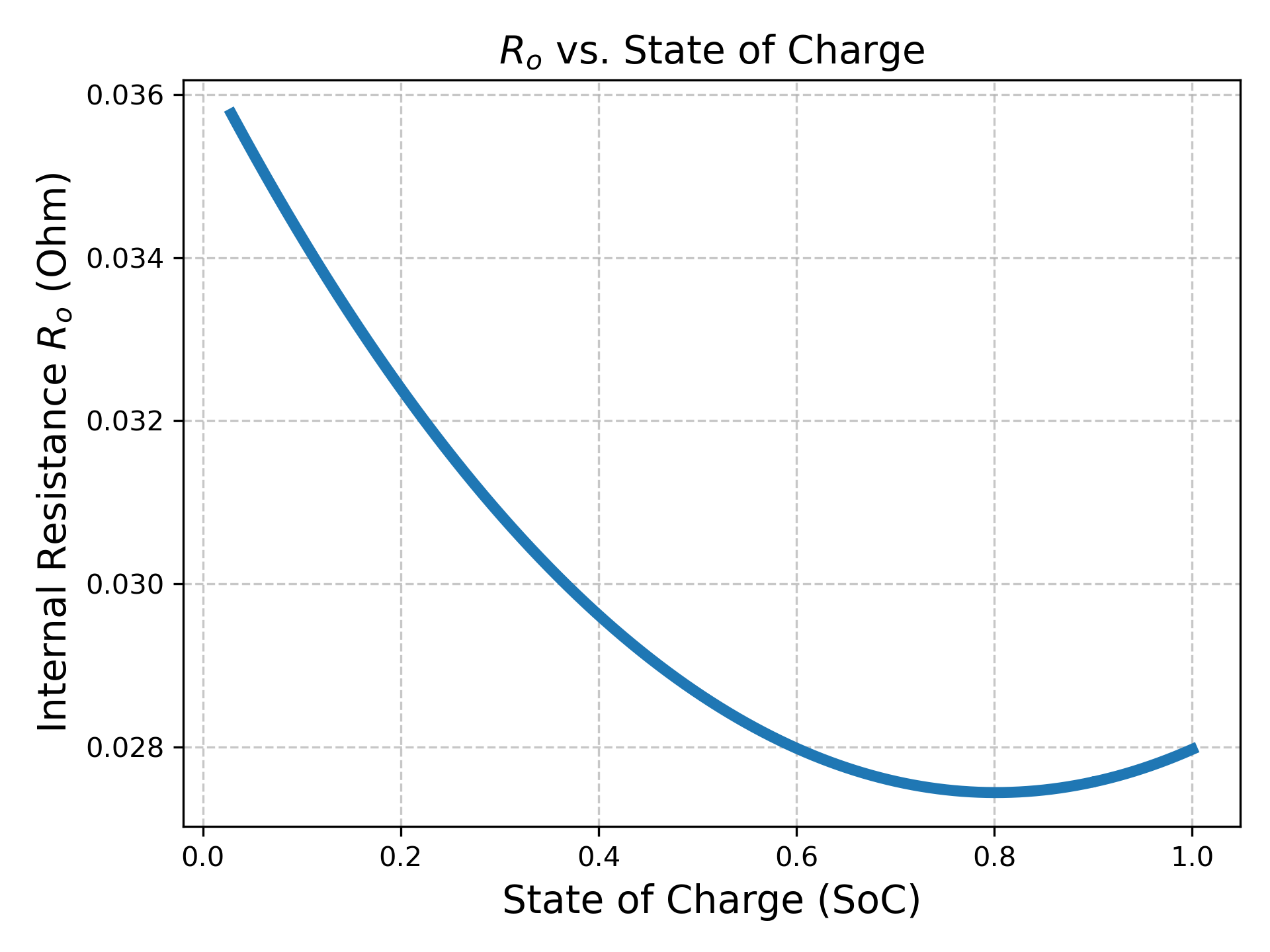}
    \caption{Plot showing fitted $R_o$ vs SoC for a single cell. It is observed that the cell's resistance reduces with increasing SoC up to a certain point and marginally increases at high SoC.}
    \label{fig:Ro_vs_SoC}
\end{figure}

Because the ECM model equations are non-convex in the desired parameters, we use a Genetic Algorithm (GA) to fit the model. Other non-linear search methods such as Particle Swarm Optimization (PSO), Neural Networks (NN), or other gradient-based methods may be used. We elect the GA due to its computational efficiency and reliability when compared to PSO. Additionally, a NN may be extraneous for a battery ECM with well-defined structure and may suffer from being stuck at saddle-points, making the module less robust. From the HPPC data, we observe the well-known phenomenon that the voltage drop for a given current $I(t)$ is SoC dependent. Consequently, the resistance $R_o$ is captured using the following relation:
\begin{equation}
    R_o(\bar{S}(t)) = B_{R_o}e^{C_{R_o}\bar{S}(t)} + A_{R_o}e^{\bar{S}(t)}
\end{equation}

where $A_{R_o}$, $B_{R_o}$, and $C_{R_o}$ are parameters that are learned by the GA. The rationale for 
the given relation can be explained visually (see Fig. \ref{fig:Ro_vs_SoC}). Typically, the resistance of the battery cell has been observed to follow the shape in Fig. \ref{fig:Ro_vs_SoC} and the proposed relation offers a way to parameterize this curve for all SoC of any cell, given its resistance follows a similar shape. $C_{R_o}$ is constrained to be non-positive, enabling the model to capture the steepness (or rate) of the decaying portion of the curve, $B_{R_o}$ controls the 'stretching' of the curve, and the second exponential is used to capture the observed rise in SoC after the initial decrease with increasing SoC. The HPPC and capacity test experimental data (at a 1 second resolution) is leveraged by the identification algorithm.

\begin{figure}
    \centering
    \includegraphics[width=7cm]{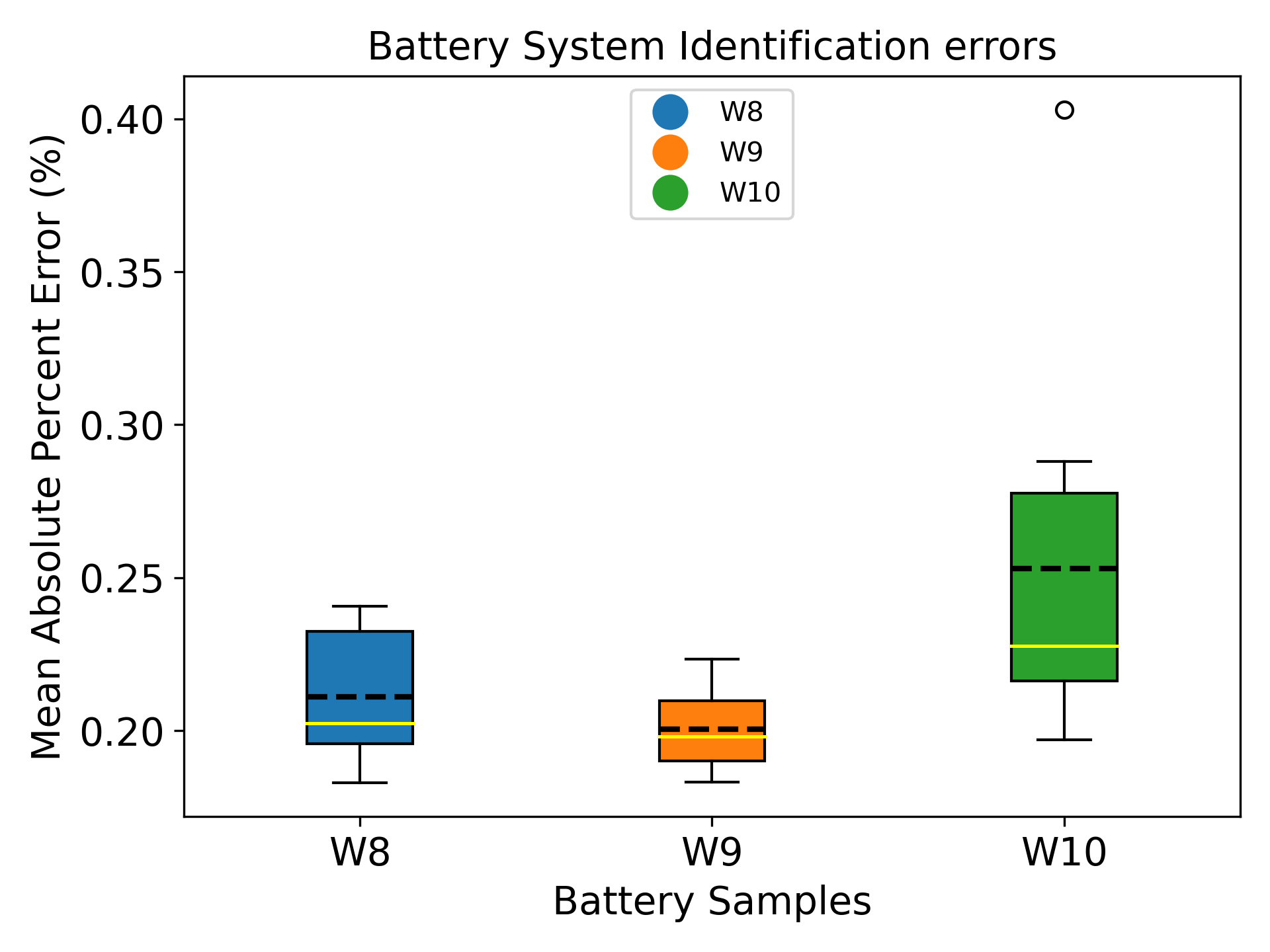}
    \caption{The identification algorithm is run on each cell for 10 trials. In all cases the errors are below 0.5\%. The black dashed lines are the means; the yellow lines are medians. The sample names are retained from the data source.}
    \label{fig:batt_iden_errorbars}
\end{figure}

To limit the algorithm's search space (for quicker convergence), a range for $R_o$ is roughly estimated from experimental data by calculating the instantaneous voltage drops experienced by the cell. Once a range is obtained, the algorithm finds the best joint set of parameters to optimize the objective (or fitness function). The experimental data is fed through the system identification module to obtain a parameter vector $\Theta = [R_1,C_1,R_2,C_2,A_{R_o},B_{R_o},C_{R_o}]^T$. Moreover, we note that a good fit can still be obtained with only a subset of the data. The fitness function is defined as the negative RMSE of the ECM times $\alpha_{fit}$:
\begin{equation}
    J(\Theta) = -\alpha_{fit}\sqrt{\sum_{i=1}^{n}\frac{1}{n}\left(V_{data, i} - V(\Theta)_i\right)^{2}}
\end{equation}
where  $V_{data, i}$ is a sample from the experimental voltage and $V(\Theta)_i$ is a corresponding voltage sample from the ECM model. $n$ is the number of measured data points. We introduce a hyperparameter $\alpha_{fit}$ to increase the fitness function's variance over the search space, encouraging exploration; we choose $\alpha_{fit} = 10$ and find it works well across all cells (see Fig. \ref{fig:batt_iden_errorbars}).

During fitting, there is a consistent bias error on the predicted voltages across all SoC due to a bias in the voltage measurement. ``Pseudo" OCV is typically measured by loading the cell with a small DC signal, usually C/20. This is bound to underestimate the OCV because: (1) the loading device and circuit wires induce voltage drops because of their own internal resistance, and (2) it is impossible to load the cell without a voltage drop. As a result of this, we introduce a novel \textit{open circuit voltage correction (or OCV correction)} scheme.
The steps for OCV correction are as follows:

\begin{enumerate}
    \item Use the initially measured OCV to fit a second-order ECM for the cell and save the learned parameters. 
    \item Using the ECM model with parameters learned from step (1), simulate the voltage response to 
    learn the quadratic bias correction function (Equation \eqref{equ:ocv_corr}), which is a function of the original measured OCV, to produce a corrected OCV that eliminates the bias error.
    \begin{equation}\label{equ:ocv_corr}
        OCV := a \cdot (OCV)^{2} + b
    \end{equation}
    \item Finally, run step (1) again using the learned OCV correction function to fit a new second-order ECM, which is the final cell model. All OCVs used in the final battery model are the corrected OCVs.
\end{enumerate}

\begin{figure}
    \centering
    \label{OCV_corr}
    \includegraphics[width=7cm]{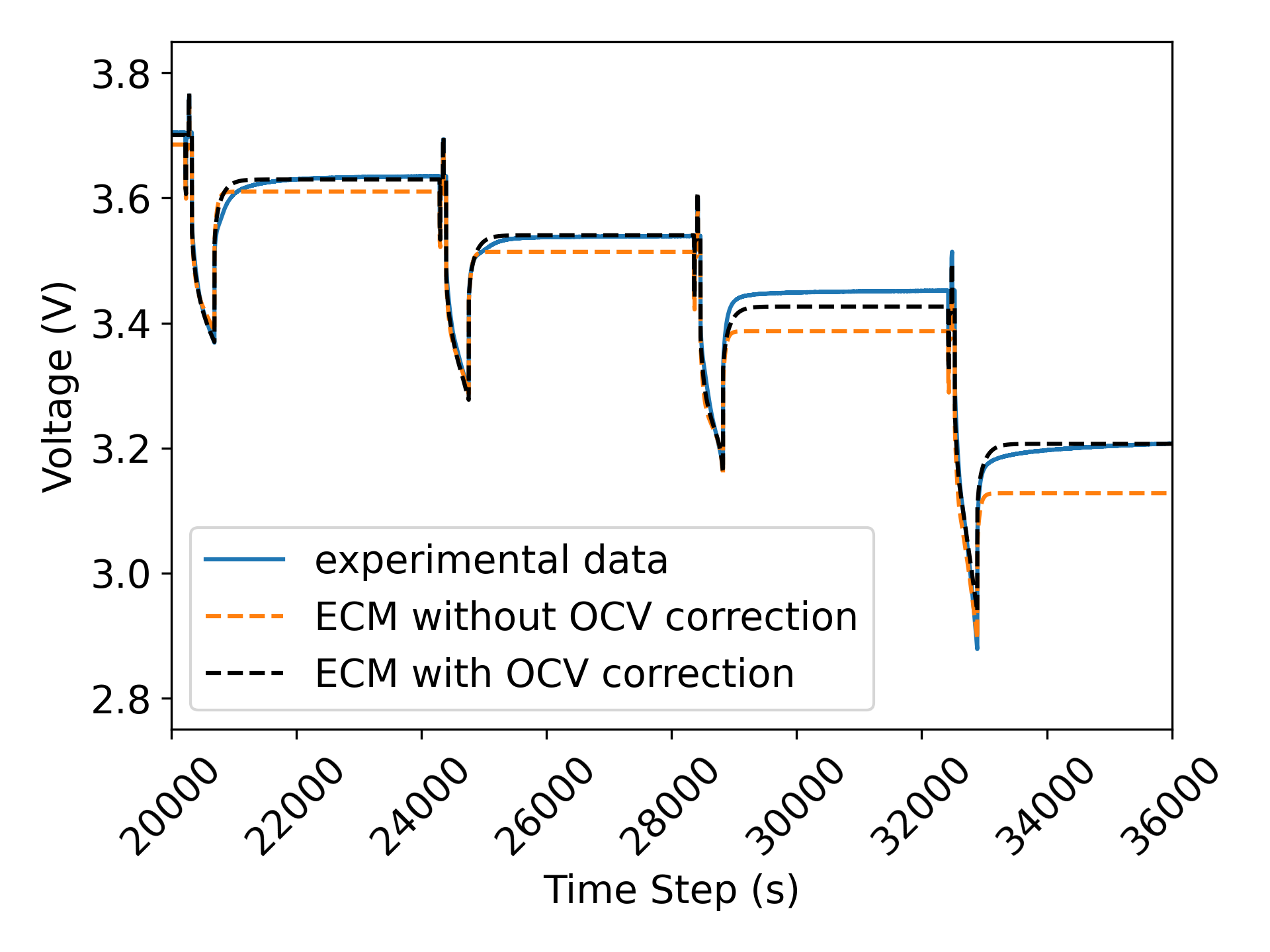}
    \caption{Plot showing the effect of OCV correction on model accuracy.}
\end{figure} 

The learned cell parameters are passed into the battery module that aggregates the cells to a pack with the desired topology. It is assumed that the cells are identical and balanced. The battery cell to pack models are obtained through impedance scaling, similar to approaches in \cite{ko_2019, abbasi_2021}. 

\begin{figure}
    \centering
    \includegraphics[width=7cm]{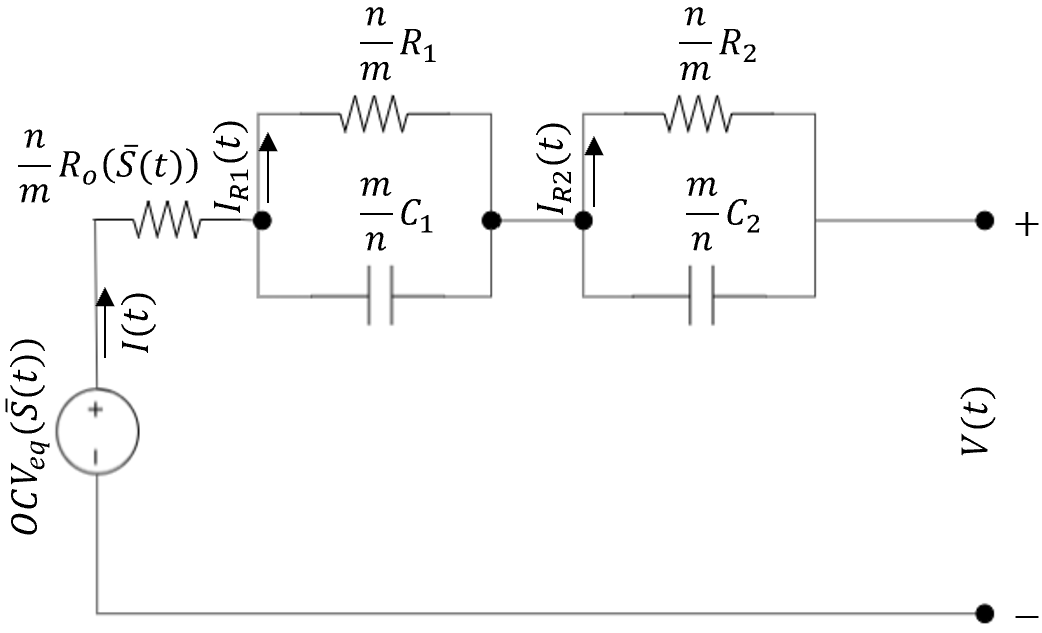}
    \caption{Second order ECM (Thevenin) circuit diagram for the pack}
    \label{fig:ecm_eq} 
\end{figure}

We briefly discusses how the battery pack model shown in Fig. \ref{fig:ecm_eq} is obtained. The impedance of a parallel RC pair is $R/(1+jRC\omega)$, which can be rewritten as  $1/(1/R+jC\omega)$. For a group of $n$ identical series impedances, the equivalent impedance for the series connected RC pairs is $1/(1/nR+jC\omega/n)$ , from which we infer an equivalent resistance and capacitance, $R_{eq,s}=nR$ and $C_{eq,s}=C/n$  respectively. Similarly, for a group of $m$ identical parallel impedances, the equivalent impedance is obtained using Ohm’s law as  $1/(m/R+mjC\omega)$ , which yields an equivalent resistance and capacitance, $R_{eq,p}=R/m $ and $C_{eq,p}=mC$.

The derived pack ECM from an initial cell model is shown in Fig. \ref{fig:ecm_eq}. $n$ represents the number of cells in series and $m$ is the number of modules in parallel. $OCV_{eq}$ is the equivalent OCV ($OCV_{eq} = n \times OCV$).

\subsection{Battery Aging} 
\begin{figure}
    \centering
    \includegraphics[width=7cm]{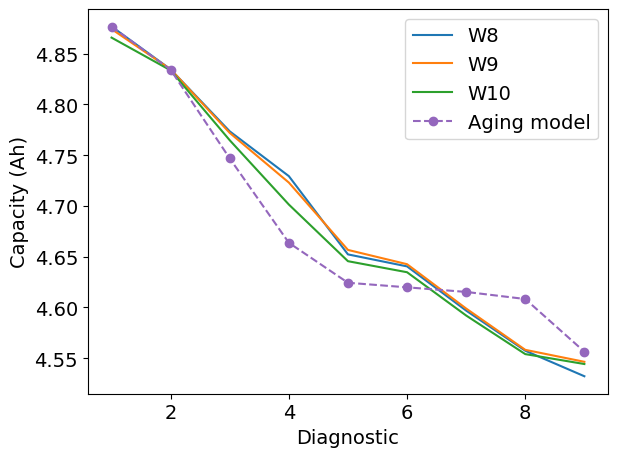}
    \caption{Plots of experimental data of 3 NMC cells and aging model. The horizontal axis diagnostic number, representing a point at which a new capacity test was carried out.}
    \label{fig:aging_model}
\end{figure}
As displayed in Fig. \ref{fig:sim_frame}, the battery aging module is distinct from the battery module. This design choice allows one to adopt and test multiple degradation models for different battery types and chemistry. It also aids easy model re-calibration and future model improvements. During each simulation time step, the aging module updates the battery module's state-of-health.
	 
The aging model used in this paper is a semi-empirical model described in \cite{schmalstieg_2014}. The reader is referred to the cited paper for a more detailed treatment of the aging model. The aging is typically captured by \textit{calendar} and \textit{cycle} aging.
\subsubsection{Calendar aging} The calendar aging of the battery occurs continuously, even when at rest. It has been shown to be temperature and voltage dependent. The temperature dependance is captured by the Arrenhius relation:
\begin{equation}
    \alpha_{T}(T) = a_1e^{\frac{-E_A}{RT}}
\end{equation} and the voltage dependence is represented as
\begin{equation}
    \alpha_V(V) = a_1V + a_2
\end{equation},
where $E_A$ is the activation energy, $R$ is the gas constant and $T$ is the thermodynamic temperature \cite{schmalstieg_2014}. $\alpha_V(V)$ and $\alpha_T(T)$ are combined in \cite{schmalstieg_2014} to obtain $\alpha_{cal}(T, V)$ and then the relation in Equation \ref{cal_aging} is used to calculated calendar aging. Parameters $a_1$ and $a_2$ are from \cite{schmalstieg_2014}.
\begin{equation}
   Q_{lost, cal} = \alpha_{cal}(T, V) t^{0.75} 
   \label{cal_aging}
\end{equation}

\subsubsection{Cycle aging} The cycle aging factor $\beta_{cap}$ is learned from experimental data and depends on the average discharge voltage and the change change in discharge depth \cite{schmalstieg_2014}. The function comprises the cycle aging factor $\beta_{cap}$ and the total charge throughput $Q(t)$ in amp-hours (Ah) at time $t$.

\begin{equation}
    Q_{lost, cyc} = \beta_{cap}\sqrt{Q}
    \label{cyc_aging}
\end{equation}
The capacity update function obtained from equations \eqref{cal_aging} and \eqref{cyc_aging} is
\begin{equation}
    Q_{cap}(t + \Delta t) = Q_{cap}(t) - Q_{lost, cal}(t) - Q_{lost, cyc}(t)
\end{equation}

 Parameters for data-driven models are usually unique to a specific cell, so parameters must be adjusted to fit a different cell with the same chemistry. We scale the aging factors in order to achieve a comparable per-cycle loss-of-life of the cells modelled in this paper\cite{pozzato_2022}. The aging plot is generated by using the same experimental current profile on the battery module; see Fig. \ref{fig:aging_model}. Each cell is subjected to a charging protocol at a specified C-rate and each diagnostic represents a capacity test after a specific number of cycles \cite{pozzato_2022}.
 
\subsection{Transformer thermal dynamics}
Arguably the most prominent transformer thermal model was introduced in \cite{swift_2001a, swift_2001}, based on a second order lumped capacitance thermal model that defines the temperature changes in the oil and hotspot because of iron loss and the copper loss in the transformer. The thermal dynamics equations are given by:
\begin{equation}
    \frac{\partial\theta_o(t)}{\partial t} = \frac{-1}{\tau_o}(\theta_o(t) - \theta_a(t))^{\frac{1}{n}} + \frac{\Delta\theta_{or}^{\frac{1}{n}}}{\tau_o}\left(\frac{K(t)^2R + 1}{R + 1}\right)
\end{equation}
\begin{equation}
    \frac{\partial\theta_h(t)}{\partial t} = \frac{-1}{\tau_h}(\theta_h(t) - \theta_a(t))^{\frac{1}{m}} + \frac{\Delta\theta_{hr}^{\frac{1}{m}}}{\tau_h}K(t)^2
\end{equation}

 The thermal state is propagated at each time step using the Euler method. $\theta_h$ is the hot spot temperature of the transformer, $\theta_o$ is the top oil temperature, and $\theta_a$ is the ambient temperature. $\tau_o$ is the top-oil time constant and $\tau_h$ is the hotspot time constant. $K(t)$ is the ratio the of current load on the transformer at time $t$ to the rated load of the transformer, $R$ is the ratio of the copper loss to the iron loss at the rated load. $\Delta\theta_{or}$ is the rated change in top-oil temperature (the change in top oil temperature at the rated load) and $\Delta\theta_{hr}$ is the rated change in hot spot temperature. $m$ and $n$ are constants based on the expected cooling mode of the transformer. The recommended values are included in the IEEEC57-91-2011 guide for loading oil-immersed transformers \cite{na_website_ndg}.

\subsection{Solar}
The solar module does not explicitly model complex dynamics. It ingests configuration files that determine the system capacity and location. Solar power output is calculated using irradiance data from the NREL US solar irradiance database \cite{na_website_ndi}. The output of the solar module is given by:
\begin{equation}
    P_{solar}(t) = \min{(P_{rated}, G_{irr}(t)A_{panel}\eta_{s}(t))}
\end{equation}
where $P_{rated}$ is the rated capacity of the solar PV, $G_{irr}(t)$ is the Global Horizontal Irradiance (GHI) at time $t$, $A_{panel}$ is the surface area of the panel and $\eta_{s}(t)$ is the efficiency.

\subsection{Charging station (EVSE)}
The charging station (or EVSE) module produces a load with a power factor parameter that determines its reactive load contribution, if any. It also retains all information of all power injection at its grid node/bus. It is initialized with its location, capacity, and efficiency. The EVSE module ingests the battery, solar and controller modules to which it is assigned. The power equations for the charging station are
\begin{equation}
    P_{grid} = P_{ev}\frac{1}{\eta_{evse}}
\end{equation}
\begin{equation}
    Q_{grid} = P_{ev} \tan{\left(\arccos{(pf)}\right)}
\end{equation}
with $pf$ as the power factor, $P_{grid}$ and $Q_{grid}$ are the real and reactive loads, and $\eta_{evse}$ is the efficiency.

\subsection{Power system}
\textit{EV-EcoSim} interfaces with GridLAB-D \cite{chassin_2008}, an open-source power systems modelling and simulation environment, to simulate the power distribution system. GridLAB-D is a simulator for running three-phase unbalanced quasi-static timeseries power flow calculations. The interface between \textit{EV-EcoSim} allows simulation variables to be passed between the two environments at each timestep in the simulation. At each timestep, the real and reactive power injections at all load buses are calculated by \textit{EV-EcoSim} modules and sent to GridLAB-D before running the power flow calculation. GridLAB-D solves the power flow problem using a Newton-Raphson solver using the current injection method \cite{chassin_2008}. Power flow solution values (voltage magnitudes and phases at load buses) can then be passed back to \textit{EV-EcoSim}, if they are used by the simulation or controller, or saved to file for post-processing and analysis. A variety of different distribution system feeder models can be simulated in GridLAB-D, such as standard IEEE test systems \cite{kersting1991radial}, synthetic taxonomy feeders \cite{schneider2009taxonomy}, or real-world feeder models. The \textit{EV-EcoSim} framework populates these network models with time-varying load profiles and DER locations. The spot loads from the original feeder models can be used to inform the placement of resources and development of time-varying load profiles.

\subsection{Controller}\label{controller}
The controller is a user-defined module that decides how controllable DERs are leveraged. A user can specify the controller to be of any form. It can be logic-based or optimization-based, with any objective, given it produces the relevant control variables at each time step. This flexibility can be useful for controller design. In this paper, we use an optimization based controller that solves a mixed-integer linear program (MILP). This is termed the \textit{linearized circuit model} because it considers the battery voltage but ignores the Joule heating loss in order to preserve convexity. Note however, the true states of the modules are evolved using the higher fidelity models described in this paper. The unavoidable asymmetry between controllers and real systems is a desirable effect we wish to capture as it better mirrors reality. 

The controller can work in either a receding horizon control (RHC) fashion or as an open-loop/one-shot optimization. In the RHC mode, only the first computed action is taken before all the states are updated. Then the next optimization problem is solved at the next time step, with the current state becoming the initial state for the next problem and the horizon shifted forward one step; this can be time-intensive. For the one-shot scheme, a single optimization problem is solved for the entire horizon and requires less computation than RHC. The RHC mode mimics the real-time operation of the system, while the one-shot mode is suitable for planning. The one-shot approach is computed offline while RHC is an online control policy that can account better for model and environment uncertainty.

\paragraph{Decision variables}
The main decision variables are $I(t)$, $I_{solar}(t)$, $I_{ev}(t)$, and $I_{grid}(t)$ as described in Tab. \ref{tab:table1}. 

\paragraph{Optimization problem} For the case study discussed Section \ref{casestudy}, the objective is to minimize the EVSE's overall electricity cost. The problem is described below as:

\begin{table}
\caption{Description of controller optimization variables\label{tab:table1}}
\centering
\begin{tabular}{|c|c|c|}
\hline
Variable & Description & Sign\\
\hline
$I_{solar}(t)$ & Current from solar to battery & (+)\\
\hline
$I_{grid}(t)$ & Current from grid to battery & (+)\\
\hline
$I_{max}$ & Max allowable battery current & (+)\\
\hline
$I_{ev}(t)$ & Current from battery for EV charging & (+)\\
\hline
$I(t)$ & Current flow in or out of battery & (+)\\
\hline
$P(t)$ & Power flow in or out of battery & (+, -)\\
\hline
$P_{ev}(t)$ & Power from battery to EV & (-)\\
\hline
$\bar{S}(t)$ & Battery state of charge & (+)\\
\hline
$S_{ev}(t)$ & Power from solar to EV & (+)\\
\hline
$L_{ev}(t)$ & EV charging load (kW) & (+)\\
\hline
\end{tabular}
\end{table}

\begin{align}
\allowdisplaybreaks
& \underset{I_{solar}, I_{ev}, I, I_{grid}}{\text{minimize}}
& & \lambda_{elec} \\
& \text{subject to}
& & \label{eq11}\bar{S}(t=0) = \bar{S}_{initial}\\
& & & \label{eq12} \bar{S}(t) = \bar{S}(t - \Delta t) + \frac{I(t)\Delta t}{Q(t)}, t > 0\\
& & & \label{eq13}
    y_{solar}, y_{grid}, y_{ev} \in {0, 1}\\
& & & \label{eq14}
    I_{solar}(t) \leq I_{max} y_{solar}\\
& & & \label{eq15}
    I_{grid}(t) \leq I_{max} y_{grid}\\
& & & \label{eq16}
    I_{ev}(t) \leq I_{max} y_{ev}\\
& & & \label{eq17}
    I_{ev}, I_{solar}, I_{grid} \geq 0\\
& & & \label{eq18}
    y_{ev} + y_{grid} \leq 1\\
& & & \label{eq19}
    y_{ev} + y_{solar} \leq 1\\
& & & \label{eq20}
    P_{ev}(t) = -I_{ev}(t)V(t-\Delta t)\\
& & & \label{eq21}
    P_{grid}(t) = I_{grid}(t)V(t-\Delta t)\\
& & & \label{eq22}
    P_{solar}(t) = I_{solar}(t)V(t-\Delta t)\\
& & & \label{eq23}
    P(t) = P_{ev}(t) + P_{solar}(t) + P_{grid}(t)\\
& & & \label{eq24}
    L_{ev}(t) + P_{ev}(t) - S_{ev}(t) \geq 0
\end{align}
$\lambda_{elec}$ is the total electricity cost and will be properly defined in Section \ref{casestudy} below. The controller constraints are defined w.r.t the battery, solar, and grid resources. The constraints are necessary to ensure the controller is working within the space of physically realistic solutions.

Equations \eqref{eq11}-\eqref{eq19} ensure the control actions respect battery physics. Equations \eqref{eq11}-\eqref{eq12} define the SoC evolution via coulomb counting. Equation \eqref{eq17} defines binary variables $y_{solar}$, $y_{grid}$, and $y_{ev}$ that control whether the battery is charging from solar PV, the grid,delivering power to the EVSE, or doing nothing (its value will be zero). Equations \eqref{eq14}-\eqref{eq19} define the currents flowing into and out of the battery and Equations \eqref{eq18}-\eqref{eq19} ensure that the controller does not consider solutions where the battery is discharging and charging at the same time. Equations \eqref{eq20}-\eqref{eq23} define the total power injected into or delivered by the battery and Equation \eqref{eq24} restricts the battery from discharging to the grid.

\section{Case study}\label{casestudy}
We demonstrate the utility of \textit{EV-EcoSim} via a case study. We examine the sizing of a battery for a site in California, USA, with a fixed 80 kW capacity solar PV installed and load profiles for the site. Solar PV is allowed to net-meter at the time-of-use (TOU) rate given to the EVSE operator. We use the Levelized Cost of Energy (LCOE) as a metric for comparing configurations across the different scenarios. We assume a perfect forecast of the EV load and insolation as we are solving a planning problem focused on sizing a DER system for different utilization scenarios.

The levelized cost of electricity is the estimated revenue or total net expenditure required to build and operate an energy system over a specified cost recovery period \cite{useia_2022}. In this text, we use the more general phrase “levelized cost of energy,” to fit the more common “levelized cost of electricity” and “levelized cost of storage (LCOS)” under one umbrella. This is done to conveniently ascribe a combined lifetime value to the system (solar + battery) rather than each component separately. The recovery period for this study is the expected operational lifetime of the energy asset.

The levelized cost of energy or LCOE of a solar PV system is equivalent to its levelized cost of electricity. We define the LCOE for the battery system as the total lifetime net expenditure per unit energy over its operational lifetime.  For the LCOE of solar calculation, we use the NREL Comparative PV LCOE Calculator \cite{na_website_ndh}, which takes specified values to produce LCOE estimates. Tab. \ref{tab:PV_params} shows the set of parameters used to calculate the LCOE for the solar system in this study.

\begin{table}
\caption{Parameters for solar LCOE estimation.}
\centering
\label{tab:PV_params}
\begin{tabular}{|c|c|}
\hline
Parameter & Value\\
\hline
Cell Technology & Mono-Si\\
\hline
Package type & Glass-polymer\\
\hline
System type & Roof-mounted commercial scale\\
\hline
Location & USA (CA)\\
\hline
Inverter loading ratio & 1.3\\
\hline
\end{tabular}
\end{table}
The baseline LCOE for PV is \$0.067/kWh. We used a fixed capital cost of \$345/kWh for the battery, from NREL \cite{cole_2021}. We obtain the LCOE for the battery with the following steps:

\begin{enumerate}
    \item  Select a fixed capital cost per energy capacity.
    \item Calculate the expected life of the battery post-simulation. It is calculated from the amount of degradation/aging experienced over the simulation time frame:
    \begin{equation}
        L_{exp} = \frac{0.2}{Q_{lost}}N_{sim},
    \end{equation} 
    where the end of life of the battery is when the battery capacity $Q(t)$ is at 80 percent of its nameplate capacity $Q_{cap}$. $Q_{lost}$ is the portion of the battery’s nominal capacity that was lost during the period of simulation. $N_{sim}$ is the number of days simulated. Therefore $L_{exp}$ is the expected life of the battery (in days) if the aging path continues as in the simulated days.
    \item Calculate the expected energy throughput of the battery over its expected lifetime. This is obtained post-simulation and described by the equation: 
    \begin{equation}
        E_{exp} = E_{daily}L_{exp}
    \end{equation} 
    where $E_{daily}$ is the average daily energy throughput.
    \item Calculate the expected battery aging cost (\$). In this work, we assume all costs due to aging incurs a cost proportional to the capital cost for capacity lost. 
    \begin{equation}
        \lambda_{aging} = \lambda_{capital} \frac{Q_{lost}}{0.2} \frac{L_{exp}}{N_{sim}}
    \end{equation}
    The equation above simplifies into the original capital cost of the battery, implying that the expected aging cost over its expected operational lifetime is equal to the original capital cost, which is reasonable. $\lambda_{aging}$ is the aging cost and $\lambda_{capital}$ is the capital cost.
    \item Calculate the total cost (normalized) over its expected life, which is the levelized cost of energy. It is cost for each unit of energy flowing through the battery during its operational life (\$/kWh):
    \begin{equation}
        \lambda_{LCOE} = \frac{\lambda_{aging} + \lambda_{capital}}{E_{exp}}
    \end{equation}
\end{enumerate}

\subsection{Scenarios and configuration}
We investigate the impacts and economic value of different BESS systems for a given set of scenarios. The scenario is defined by the expected utilization level and environment. The EVSE controller objective is to minimize the electricity cost.

In this case study, we consider two degrees of freedom: the maximum allowable C-rate of the battery and its energy capacity. Five C-rates ($C_{sim}=\{0.1,0.2,0.5,1.0,2.0\}$) and five capacities ($E_{sim}=\{50,100,200,400,800\}$kWh) are considered, for a total of 25 simulations per month. 

The station level EV charging load profiles are generated using the SPEECh model \cite{powell_2022} at varying levels of EV penetration (see Tab. \ref{tab:table3}). The power a charging station can deliver is capped by its predetermined capacity. 

\begin{table}
\caption{Summary load characteristics for simulated load scenario\label{tab:table3}}
\centering
\begin{tabular}{|c|c|c|}
\hline
 No. EVs & Peak load (kW) & Average load (kW)\\
\hline
400 & 172 & 22\\
\hline
800 & 354 & 45\\
\hline
1600 & 424 & 91\\
\hline
3200 & 608 & 178\\
\hline
\end{tabular}
\end{table}

For the distribution network, we use the IEEE 123 bus network \cite{kersting1991radial}. Data from the Pecan Street database \cite{parson2015dataport} was used to populate time-varying residential building loads within the power network. The magnitude of the residential load (e.g. number of homes) at each node was sized relative to the specified spot loads. The charging station transformers were all sized at 75 kVA, for comparing the impacts of different systems design on transformers.

\subsection{Cost function}
 We use the PGE BEV2-S electricity bill for business EVSE in the objective \cite{PGE_BEV_SCH}. The rate structure includes a TOU rate and an additional subscription charge for EVSE operators to purchase a maximum penalty-free average power that can be consumed within a 15-minute interval. The subscription charges are sold based on a block system. If a customer exceeds their allowable maximum power for any 15-minute window within a month, the customer is charged an overage fee (per kW) for each kW within the 15-minute window with the maximum power subscription exceedance. We express this electricity cost mathematically below.

\begin{equation}
    \label{sub_cost}
    \lambda_{sub} = \gamma_{b}p_{b}
\end{equation}
\begin{equation}
    \label{overage_cost}
    \lambda_{over} = p_{over} \underset{t}{\max}\{P_{grid}(t) - \gamma_{block}P_{all}\}_{+} \forall t \in T
\end{equation}
\begin{equation}
    \label{tou_cost}
    \lambda_{tou} = \sum_{t=0}^{T}p_{tou}(t)P_{grid}(t)
\end{equation}
\begin{equation}
    \label{tot_cost}
    \lambda_{elec} = \lambda_{tou} + \lambda_{over} + \lambda_{sub}
\end{equation}

$\lambda_{sub}$ is the subscription cost, $\gamma_{b}$ is an integer decision variable, which is the number of blocks to be purchased and $p_{b}$ is the price per block. $\lambda_{over}$ is the overage cost, $p_{over}$ is the overage fee, $P_{grid}$ is the net grid load from the EVSE, and $P_{all}$ is the power allocated per block. $p_{tou}$ is the TOU price.

\section{Results and discussion}\label{results}

\begin{table}
\setlength{\tabcolsep}{4pt}
\caption{June LCOE (USD/kWh) with 0.1 C max cycle constraint. \label{tab:june_cost_mat}}
\centering
\begin{tabular}{|c|c|c|c|c|c|c|}
\hline
No. EVs& Base & 50 & 100 & 200 & 400 & 800\\
\hline
400 & 0.2573 & -0.0508 & -0.0491 & -0.0488 & -0.0486 & -0.0406\\
\hline
800 & 0.2536 & 0.1396 & 0.1399 & 0.1402 & 0.1404 & 0.1442\\
\hline
1600 & 0.2461 & 0.2271 & 0.2172 & 0.2375 & 0.2388 & 0.1657\\
\hline
3200 & 0.2432 & 0.2639 & 0.2694	& 0.2695 & 0.2694 & 0.2229\\
\hline
\end{tabular}
\end{table}

\begin{table}
\caption{January LCOE (USD/kWh) with 0.1 C max cycle constraint.
\centering
\label{tab:jan_cost_mat}}
\begin{tabular}{|c|c|c|c|c|c|c|}
\hline
No. EVs & Base & 50 & 100 & 200 & 400 & 800\\
\hline
400 & 0.2573 & 0.2197 & 0.2198 & 0.2202 & 0.2269 & 0.2279\\
\hline
800 & 0.2536 & 0.2743 & 0.2744 & 0.2745 & 0.2777 & 0.2783\\
\hline
1600 & 0.2461 & 0.2827	& 0.2723 & 0.2929 & 0.2191 & 0.2198\\
\hline
3200 & 0.2432 & 0.3031 & 0.3032 & 0.3031 & 0.3031 & 0.2565\\
\hline
\end{tabular}
\end{table}

\begin{figure}
\centering
\includegraphics[width=8.5cm]{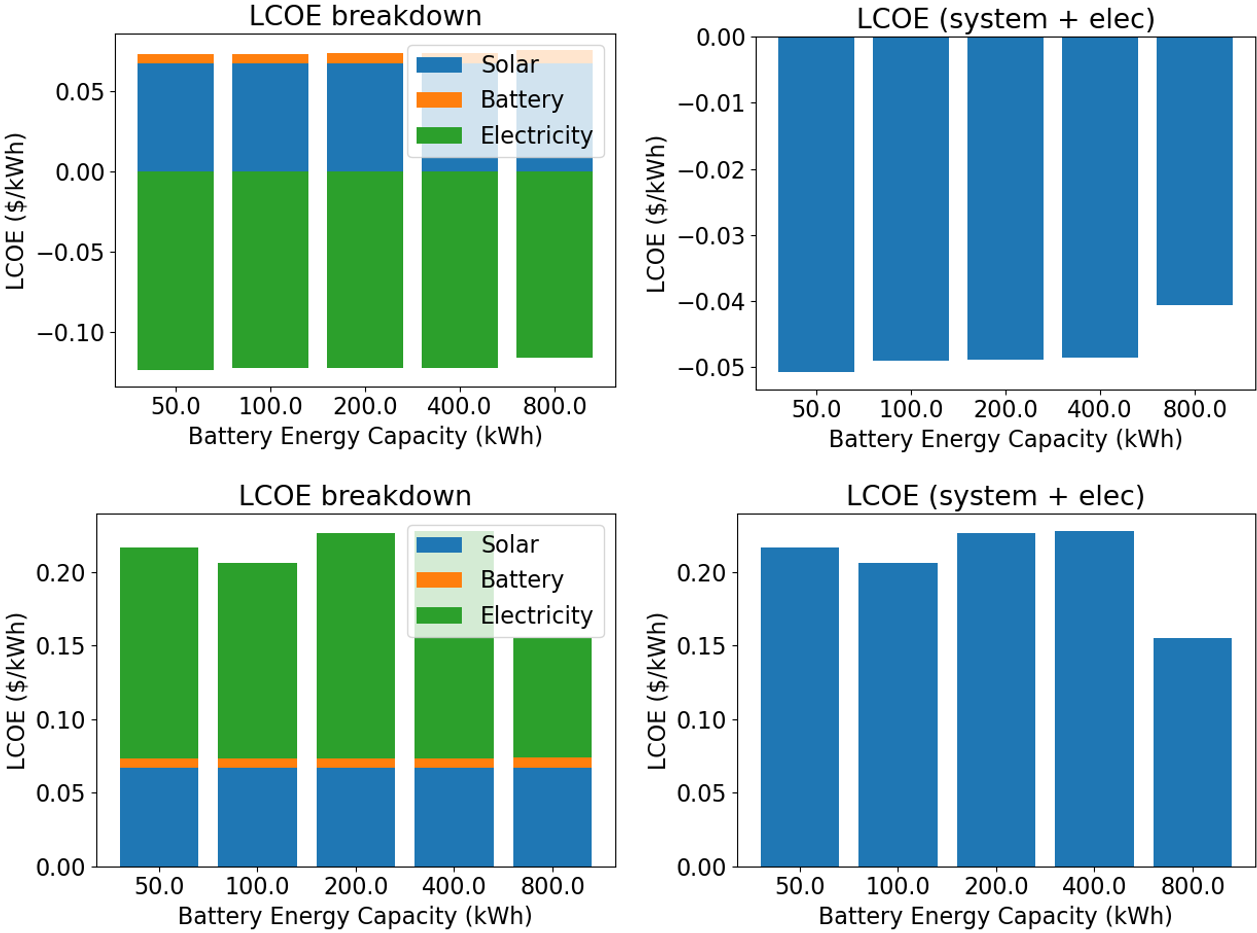}
\caption{Plots showing the LCOE for two different charger utilization scenarios in the summer month of June. Top: 400 EV population scenario, bottom: 1600 EV population scenario.}
\label{fig:bar_costs}
\end{figure}

\begin{figure}
    \centering
    \includegraphics[width=6cm]{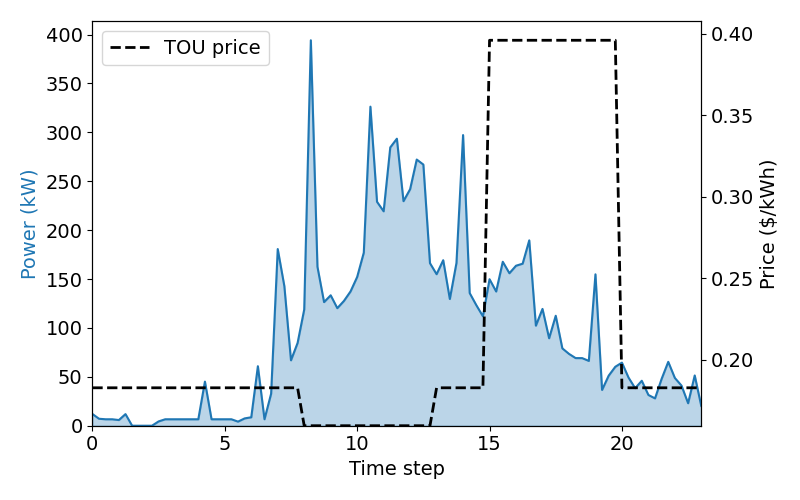}
    \caption{Load shape for the simulated aggregate EV charging profile. The TOU rates (dotted black) show that the peak loads roughly coincide with off-peak and super-off-peak prices.}
    \label{fig:load_TOU}
\end{figure}

Tab. \ref{tab:june_cost_mat} contains estimates for the LCOE (electricity + battery) in dollars per kilowatt-hour (\$/kWh) for June. In this case study, the solar capacity was not varied, and thus is fixed for all configurations in the cost comparison matrix. The baseline (no DER) is more expensive than the configurations with DER in most cases, except at the 3200 EV load scenario for which an 800 kWh battery capacity is needed to outperform the baseline. For the 400 EV load scenario, all BESS configurations outperform the base case. Due to low utilization in the 400 EV scenario, the system generates revenue by net metering excess solar energy, yielding net profits. With 400 and 800 EVs simulated to generate load profiles, respectively, there is no overall added benefit for the increased battery capacity, as can be observed in Fig. \ref{fig:bar_costs}. In the 400 EV (top) scenario in Fig. \ref{fig:bar_costs}, there is an increase in the system levelized cost (or decrease in profit)---this is because the marginal cost of the battery exceeds the marginal profit from additional capacity. However, with 1600 and 3200 EVs, respectively, the overall system cost for the simulated month reduces as the battery size is significantly increased, implying there is an added benefit of increased energy capacity---this is largely driven by the fact that higher charging loads create more opportunity for load shaping by the DER. There is a slight increase in electricity cost for the 1600 EV load scenario with 200 and 400kWh batteries in Fig. \ref{fig:bar_costs}. This is mainly driven by sub-optimal battery control signals (incurring an intrinsic cost) due to the controller's battery model error. From Tab. \ref{tab:june_cost_mat}, we report a 115.5\%, 43.1\%, 36.7\%, 8.3\%, reduction in the combined LCOE between the baseline and 800 kWh configuration, for 400, 800, 1600, and 3200 EV load scenarios, respectively. We notice a downward trend in the marginal cost savings as the scenario load increases, mainly because: (1), higher utilization levels mean there is rarely any excess solar to sell back to the grid and (2), the load peak is not coincident with peak prices,
(see Fig. \ref{fig:load_TOU}), thus inherently reducing the marginal benefit from shaping the load.

We also report the cost matrix for the month of January in Tab. \ref{tab:jan_cost_mat}. The overall cost for operating and delivering EV charging services is in general higher for the month of January than June, mainly due to lower solar irradiance levels which limits solar PV generation. We report an 11.4\%, and 10.7\% decrease in system costs (compared to baseline) for 400 and 1600 EV load scenarios respectively and in general an increase in costs for the other scenarios.

These tables could inform a planner on the relative economic benefits of multiple configurations, given an EVSE’s capacity and utilization levels. For example, for the 400 EV load scenario, the results suggest EV charging provider can offer low rates if the solar system is allowed to sell excess back to the grid---with the caveat that this benefit depends on distribution circuit's hosting capacity. There is a strong dependence of overall system economics on environmental conditions as well. For example, for the 400 EV scenario in June, adding a 50 kWh battery reduced the overall cost of energy (compared to the baseline) by more than 100\% and yielding profits. The results for January are dramatically different, owing to the dissimilar levels of irradiation during the respective seasons. It is worth noting that the overall cost of the system is not static and the expected load profile changes over time can determine if additional DER investment will be economical for a given EV charging site.

\begin{figure}
    \centering
    \includegraphics[width=8cm]{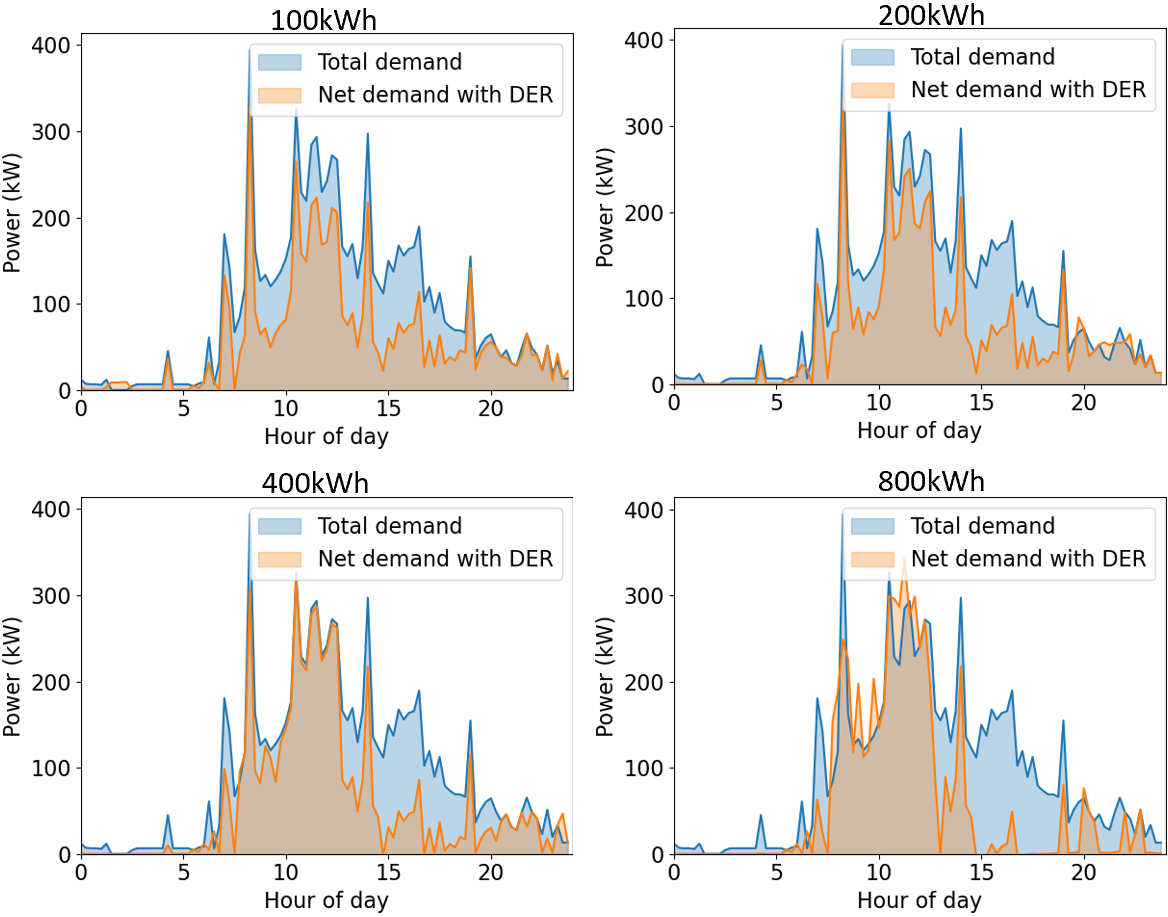}
    \caption{Plots showing initial load and reshaped load for a day in June. The blue curves are the initial electricity demand from the grid and the orange curve is the net-grid demand after optimization. Observe that the bigger 800kWh battery offsets majority of load at the peak price period.}
    \label{fig:load_shaping}
\end{figure}

The value of DERs to any system depends on their ability to arbitrage. As shown in Fig. \ref{fig:load_shaping}, the load shaping ability of the system increases as the battery energy capacity increases. However, the marginal change of the load profile reduces significantly as the capacity increases, which visually explains why increasing the battery size does always not improve overall system costs in Fig \ref{fig:bar_costs}.

\subsection{Transformer impacts}

\begin{figure}
    \centering
    \includegraphics[width=6cm]{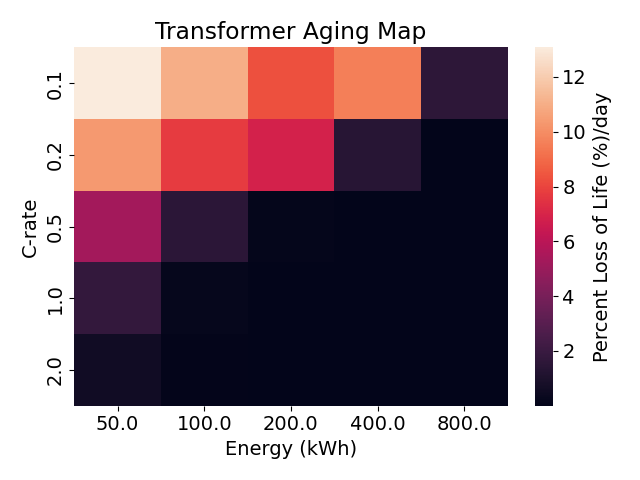}
    \caption{Transformer aging map for 1600 EV population scenario in June. Moving along this matrix, one can observe the potential impacts of lifetime operation on transformer health.}
    \label{fig:trans_aging_June}
\end{figure}

We estimate the impact of EV charging from the perspective of the distribution transformers. We show the relative impacts of the different EVSE configurations on the DCFC transformer. Observe that for a fixed BESS capacity, operating the battery at a higher C-rate in general reduces transformer rating needed for that utilization level, however sometimes it can marginally depend on controller errors. Similarly for a fixed C-rate, increasing the battery size reduces transformer rating for similar utilization levels. This is because a higher maximum discharge power allows demand spikes to be modulated. One can move along Fig. \ref{fig:trans_aging_June} to find the least cost grid feasible solution for an EVSE. For this simulation, the 800 kWh battery at a 0.5C rate has the least impact on the transformer with no accelerated aging and the transformer life being preserved beyond its normal life. Meanwhile, in the worst case, it would last only a couple days. With \textit{EV-EcoSim}, one can answer the question: \textit{what is the least-cost system design for an EVSE with a 75kVA-rated transformer}? A user could also investigate the blended costs of the transformer, battery, solar, and grid electricity over the operational lifetime of the system.

\subsection{Power system bus voltages}
Voltage violations are defined per ANSI C84.1 standard as voltages that deviate more than 5\% from the nominal. We follow a procedure for comparing a base-case to different scenarios. We develop a baseline network, which is a feeder without EV charging that does not experience any voltage violations under normal operation. Thus, we can estimate the marginal impact the EVSE would have on the network, which is shown in the results below. This type of analysis is useful to both EV charging service operators and utility companies.


\begin{table}
\caption{Percentage bus voltage violations in June for 1600EV load scenario with 50kWh battery}
\centering
\label{tab:voltage_C-rate}
\begin{tabular}{|c|c|c|c|c|c|}
\hline
Battery C-rate & 0.1 & 0.2 & 0.5 & 1 & 2\\
\hline
\% violations & 0.0204 & 0.0160 & 0.0000 & 0.0000 & 0.0000\\
\hline
\end{tabular}
\end{table}

\begin{table}
\setlength{\tabcolsep}{4pt}
\caption{Percentage bus voltage violations for June 1600 EV load scenario with C-rate 0.1 C}
\centering
\label{tab:voltage-capacity}
\begin{tabular}{|c|c|c|c|c|c|}
\hline
 Battery capacity (kWh) & 50 & 100 & 200 & 400 & 800\\
\hline
\% violations & 0.0204 & 0.0160 & 0.0067 & 0.0000 & 0.0000\\
\hline
\end{tabular}
\end{table}

Tab. \ref{tab:voltage_C-rate} and \ref{tab:voltage-capacity} show the resulting voltage violation frequency as percentages. In Tab. \ref{tab:voltage_C-rate}, we display the percentage voltage violations with varying maximum battery C-rates, while in Tab. \ref{tab:voltage-capacity}, the C-rate is fixed and the battery capacity is varied. We observe that at a fixed 0.1C rate, voltage violation frequency reduced with increased capacity, with the 400 kWh and 800 kWh batteries completely avoiding violations. The EVSE responds to prices, which are higher in the evening (see Fig. \ref{fig:load_TOU}) and uncontrollable residential loads exist within the network, which are much higher in the evening as well. Consequently, the battery is incentivized to offset the coincident EV load and residential peak load. Notice in Fig. \ref{fig:load_shaping}, the the larger batteries shave more of the evening load.

By investigating the combined economic ramifications for load scenarios, environmental conditions, and grid impacts, one can select the best battery size for a specific use-case. For instance, from the results displayed in this text, if the 1600 EV load scenario generated is the most likely, with the C-rate constraint at 0.1C, then the 800 kWh will be the best choice considering the combined cost and grid-impact.

\subsection{The cost of model fidelity}

\begin{figure*}
    \centering
    \subfloat[]{\includegraphics[width=0.31\linewidth]{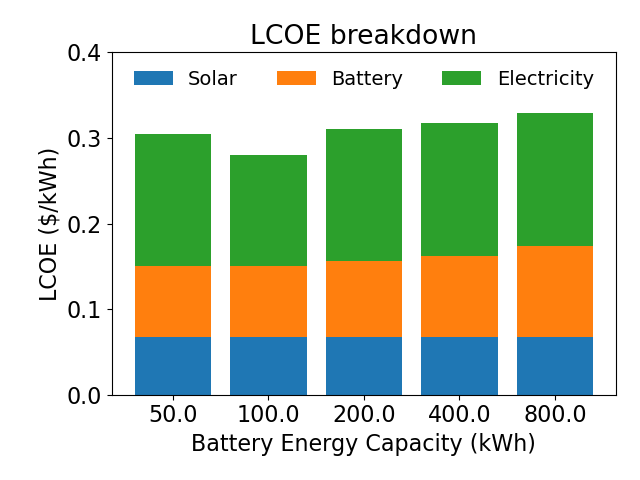}\label{fig:BC-BM}
    }
    \subfloat[]{\includegraphics[width=0.31\linewidth]{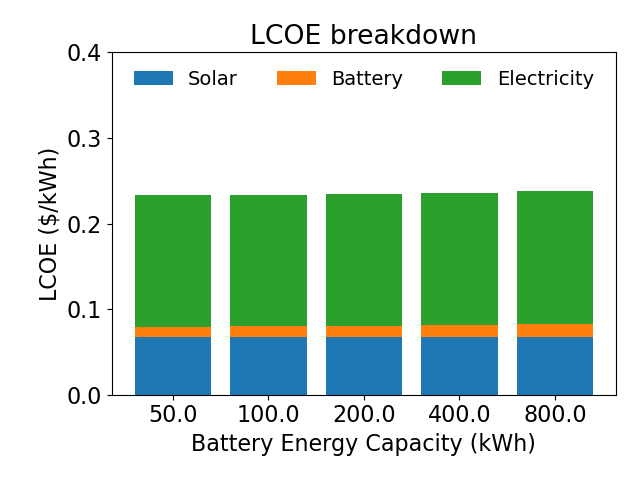}\label{fig:BC-ECM}}
    \subfloat[]{\includegraphics[width=0.31\linewidth]{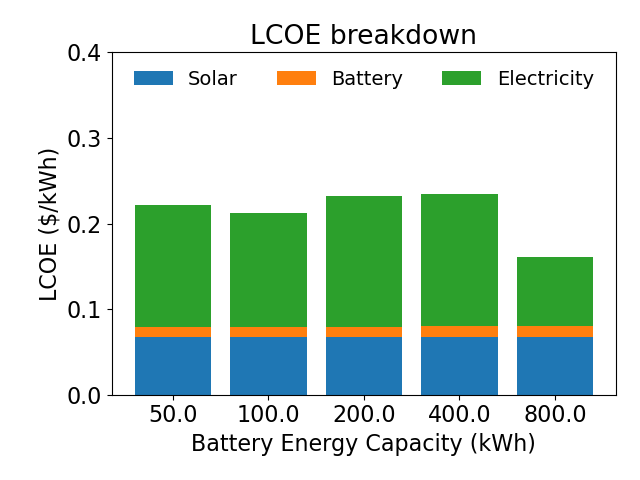}\label{fig:LC-ECM}}
    \caption{Comparing LCOE estimates for different modelling conditions with a C-rate of 0.1. (a) controller: bucket model, battery: bucket (linear) model or BM. (b) controller: bucket Model, battery: ECM with non-linear aging (in this paper). (c) controller: linearized circuit model, battery: ECM with non-linear aging (in this paper). Observe that the modelling conditions can completely alter conclusion gleaned from optimization problem.}
    \label{fig:batt-model-comp}
\end{figure*}

As previously suggested, many works do not capture the asymmetry between a controller's model and the true dynamics of the system. However, it is important to understand how model fidelity and asymmetry between a controller's model and the true dynamics of a system can affect decision-making. In this section, we consider the potential impacts of model fidelity on estimated system costs/revenue.

The charts in Fig. \ref{fig:batt-model-comp} display the different LCOE estimates under different modelling conditions. We present three modelling and simulation conditions, with increasing fidelity from left to right. Fig.~\ref{fig:BC-BM} uses a bucket model (BM) battery simulation with a BM controller, which is arguably the most common model used in storage size optimization for EV charging, Fig. \ref{fig:BC-ECM} uses a BM controller with the ECM battery model, and Fig. \ref{fig:LC-ECM} includes a linearized circuit model (described in paper Section \ref{controller}) with the ECM battery developed in this paper as well. As discussed in the text, the BM battery model does not incorporate any real physics, and the control variable is the battery power; this is a common linear model used by authors in \cite{hesse_2017a, hesse_2019}, and \cite{xu_2018}. The plots show that the differences in estimated system LCOE due to (1) controller model fidelity and (2) battery model fidelity are significant. For example, in Fig. \ref{fig:batt-model-comp}, the LCOE estimate in (c) is \$0.16/kWh while it is \$0.33/kWh and \$0.24/kWh in (a) and (b), respectively. This further buttresses the importance of higher fidelity integrated systems, which we introduce in this paper.

\begin{figure}
    \centering
    \subfloat[]{\includegraphics[width=6.9cm]{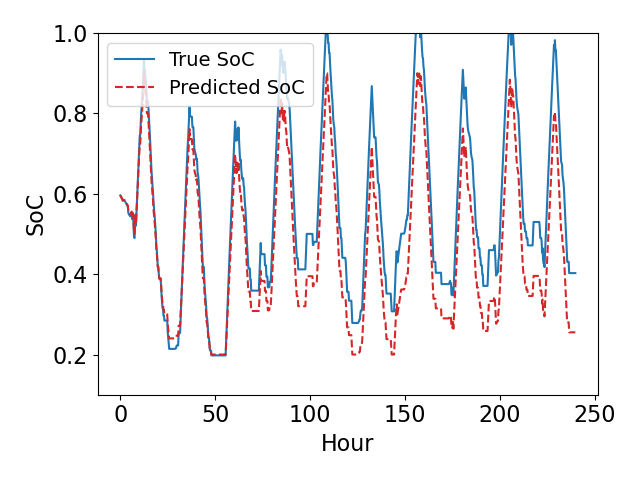}\label{fig:BM-soc_traj}}\
    \subfloat[]{\includegraphics[width=6.9cm]{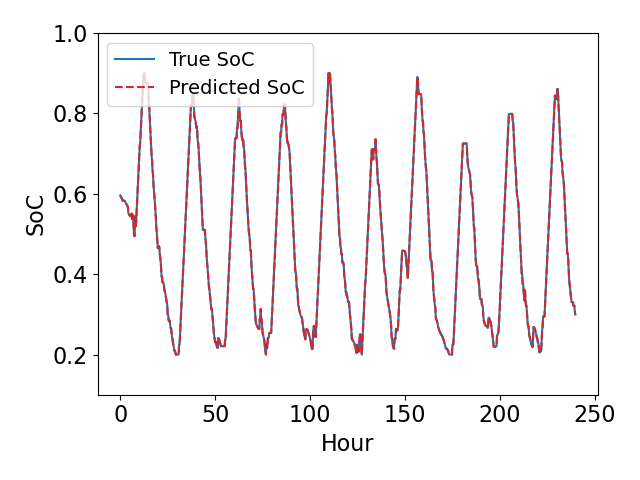}\label{fig:LECM-soc_traj}}
    
    \caption{Snippets from June simulation comparing 800kWh battery SoC trajectories for (a) bucket model and (b) linearized circuit model, with the ECM battery model (true soc). Battery C-rate is 0.1. The linearized circuit model tracks the battery's state better than the bucket model.}
    \label{fig:SoC-traj-comp}
\end{figure}

The discrepancies in the LCOE can be attributed to the fact that bucket models severely ignore the dynamics, operating voltage limits, and relaxation times of the battery. The lack of consideration for the voltage limits can significantly overestimate the power delivery capability of the battery, leading to suboptimal control trajectories, which can only be mitigated by expensively over-sizing the battery system. Plots in Fig. \ref{fig:SoC-traj-comp} show that the difference in the BM controller's predicted SoC trajectory and the battery's true SoC are significant. On the SoC trajectory, the BM controller (Fig. \ref{fig:BM-soc_traj}) has a maximum absolute and mean percent errors of 49.7\% and 23.7\% respectively, while the linearized circuit model (Fig. \ref{fig:LECM-soc_traj}) has a maximum absolute percent and mean percent errors of 0.025\% and 0.003\% respectively. The incongruence of the controller's battery state estimate and the true state of the battery explains the bucket model's inferior results when compared to the linearized circuit model. 

\section{Conclusion}\label{conclusion}
This paper introduced \textit{EV-EcoSim}, an open-source Python-based co-simulation platform that couples Electric Vehicle Charging Stations and DERs with power networks. It provides economic and grid-impact analysis to aid the planning and operation of EV charging infrastructure, including the sizing of DERs in the most grid-feasible and economic way. We demonstrated the capability of \textit{EV-EcoSim} through a case study to evaluate collocated battery storage for an EVSE.

In real-time, model-based controllers perform calculations on lower-order models due to computational constraints. Many existing studies do not consider the asymmetry between the physical system and the models many controllers would adopt, leading to irreducible errors that can change the economics of the system significantly; \textit{EV-EcoSim} addresses this. Capturing the incongruence between controller and system dynamics enables one to study the impacts of model uncertainty on system costs. This paper shows that modelling choices of the battery system can completely change the conclusion one could make from an optimization study. For example, there is an LCOE discrepancy of up to 90\% for the 800kWh battery in Fig. \ref{fig:batt-model-comp}. Further work can be done to improve the model fidelity of subsystems within the \textit{EV-EcoSim} framework to improve planning decisions.  In future work, physical test beds can be leveraged to calibrate many of the subsystems within this platform, further improving its realism.

The insights gleaned from co-simulation platforms are bound by the fidelity of underlying models, and the accuracy of the interplay between critical subsystems. Within \textit{EV-EcoSim}, we can conveniently improve internal models as higher quality data becomes available, underlying physics are better understood, and computational efficiency improves, significantly accelerating system design and the economical and equitable deployment of EV charging infrastructure.

\bibliographystyle{IEEEtran}
\bibliography{ev50}

\vfill

\end{document}